\documentclass{article}
\usepackage{spconf,amsmath,graphicx,hyperref}
\usepackage{algorithm,algorithmic}
\usepackage{multirow}
\usepackage{makecell}
\usepackage{amsmath}
\usepackage{amssymb}
\usepackage{amsfonts}

\title{PRoADS: Provably Secure and Robust Audio Diffusion Steganography with latent optimization and backward Euler Inversion}
\name{Yongpeng Yan\textsuperscript{2}, Yanan Li\textsuperscript{2}, Qiyang Xiao\textsuperscript{2}, Yanzhen Ren\textsuperscript{1,2,*}\thanks{This work is supported by the Natural Science Foundation of China (NSFC) under the grant NO.62572358,621723064,62372334. \\ * Corresponding author.  }}
\address{
\textsuperscript{1}Key Laboratory of Aerospace Information Security and Trusted Computing, Ministry of Education. \\
\textsuperscript{2}School of Cyber Science and Engineering, Wuhan University
}

\begin{document}
\ninept
\maketitle
\begin{abstract}
This paper proposes PRoADS, a provably secure and robust audio steganographic framework based on audio diffusion models. As a generative steganography scheme, PRoADS embeds secret messages into the initial noise of diffusion models via orthogonal matrix projection. To address the reconstruction errors in diffusion inversion that cause high bit error rates (BER), we introduce Latent Optimization and Backward Euler Inversion to minimize the latent reconstruction and diffusion inversion errors. Comprehensive experiments demonstrate that our scheme sustains a remarkably low BER of 0.15\% under 64 kbps MP3 compression, significantly outperforming existing methods and exhibiting strong robustness.
\end{abstract}
\section{Introduction}
\label{sec:intro}

With the development of generative artificial intelligence, the proportion of AI-generated content on the internet is continually increasing. This creates a favorable communication channel for generative steganography.
Generative steganography embeds messages during the generation of multimedia content, offering superior security and sample diversity compared to modification-based steganography\cite{wei2022generative}.
Current generative audio steganography primarily utilizes Generative Adversarial Networks (GANs)\cite{li2023coverless,chen2023imperceptible} and Flow models\cite{chen2021distribution}, but the generated content suffers from limitations concerning speech and specialized datasets\cite{zhang2025birdssong}.
Diffusion models have proven to possess powerful generative capabilities in practice, with numerous impressive works emerging in the field of audio generation \cite{liu2023audioldm, hai2024ezaudio, evans2024fast}.

Current steganographic methods based on diffusion models can be categorized into three types: generated-content-based steganography\cite{zhang2023steganography,yu2023cross}, intermediate-state-based steganography \cite{jois2024pulsar,peng2023stegaddpm,wu2024robust}, and initial-noise-embedding-based steganography \cite{wei2023generative,kim2025diffusion,yang2024gaussian,hu2024establishing}.
Among these, initial-noise-embedding-based methods have the distinct advantages that their generation process is identical to the normal generative process, and they do not require synchronized random seeds for generation. These methods obtain the reconstructed initial noise via diffusion inversion and extract the secret message from it. The differences between various steganographic schemes primarily lie in how the hidden message is mapped into the diffusion model's initial noise.
Wei \cite{wei2023generative} proposed the GSD method, which maps the secret message into the DCT coefficients of the initial noise and reconstructs Gaussian-distributed noise via IDCT.initial noise of the diffusion model.
Kim \cite{kim2025diffusion} proposed the DiffStega method, which maps the secret message to a binary center, striking a balance between distribution preservation and robustness.
Yang \cite{yang2024gaussian} proposed an interval mapping method, using a quantile function to map uniformly distributed messages into Gaussian-distributed sampling points.
Hu \cite{hu2024establishing} mapped the secret message into a standard Gaussian distribution through matrix multiplication by constructing an orthonormal matrix.
While these methods improve the robustness of the steganographic system through carefully designed mapping algorithms, they do not adequately address the latent variable reconstruction errors and diffusion inversion errors introduced during diffusion inversion. These errors impair the reconstruction accuracy of the initial noise during message extraction.

In this paper, we propose a steganography method based on latent space initial noise mapping for audio diffusion models. By incorporating latent space optimization and backward Euler inversion, our method enhances the reconstruction accuracy of initial noise, achieving robust steganography with higher precision. The key contributions are as follows:

1) We propose a generative steganography method based on audio diffusion models, which embeds secret messages into the initial noise through orthogonal matrix projection to achieve robust and provably secure steganography.

2) We introduce latent space optimization and backward Euler iteration in the process of extracting messages from stego audio, which are applied to latent reconstruction and diffusion inversion, respectively. These two approaches reduce the error in reconstructing the initial noise, thereby lowering the BER during message extraction.

3) Experiments demonstrate that our method maintains a BER of 0.15\% under 64 kbps MP3 compression, highlighting its exceptional robustness.


\begin{figure*}
    \centering
    \includegraphics[width=1\linewidth]{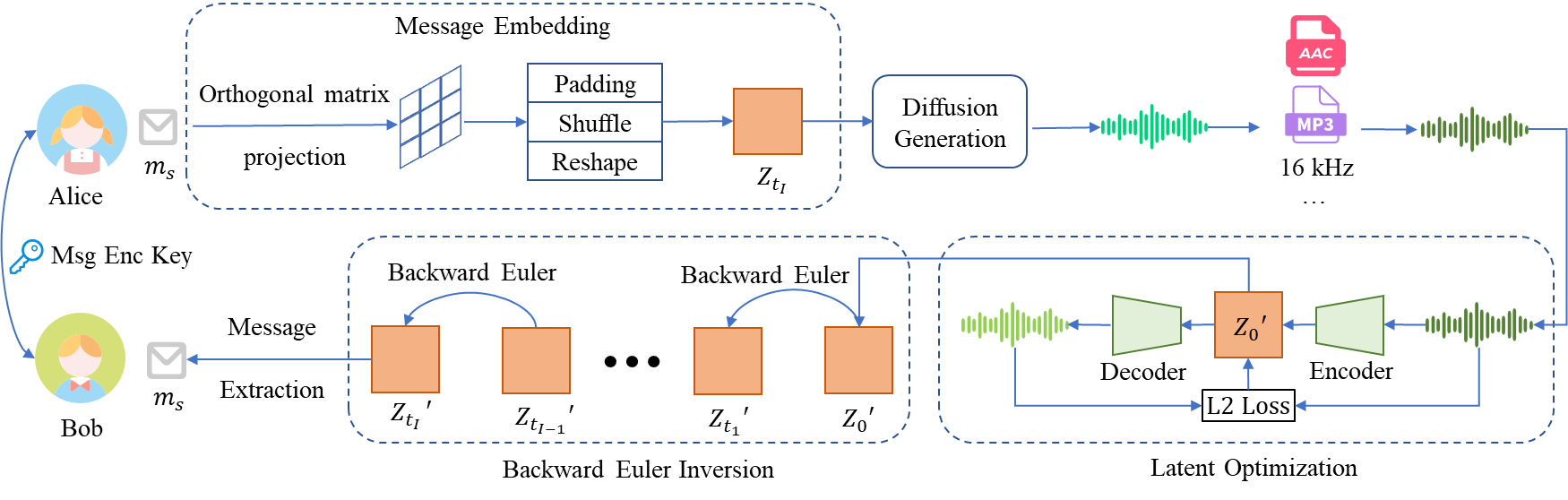}
    \caption{The overall framework of the PRoADS.}
    \label{fig:1}
\end{figure*}

\section{Preliminaries}
DPM-Solver represents a family of numerical solvers for diffusion models, with its first-order and second-order formulations being widely employed for accelerated diffusion sampling. The primary objective is to achieve computational efficiency in sampling by optimizing the diffusion ordinary differential equation (ODE) solving process. By performing a high-order Taylor expansion on the exponential integrator-based diffusion ODE framework introduced by Lu et al.\cite{lu2022dpm}, various solver forms can be derived based on the expansion order, yielding the first-order formulation (which is equivalent to DDIM sampling):
\begin{equation}
\label{eq:1}
    x_{t_i} = \frac{\sigma_{t_i}}{\sigma_{t_{i-1}}}x_{t_{i-1}}-\alpha_{t_i}(e^{-h_i}-1)x_{\theta}(x_{t_{i-1}},t_{i-1})
\end{equation}
and the second-order formulation:
\begin{equation}
\label{eq:2}
\begin{split}
    x_{t_i} &= \frac{\sigma_{t_i}}{\sigma_{t_{i-1}}}x_{t_{i-1}}-\alpha_{t_i}(e^{-h_i}-1)x_{\theta}(x_{t_{i-1}},t_{i-1}) \\
    &\quad - \frac{\alpha_{t_i}(e^{-h_i}-1)}{2r_i}(x_\theta(x_{t_{i-1}},t_{i-1})-x_{\theta}(x_{t_{i-2}},t_{i-2}))
\end{split}
\end{equation}
Specifically, diffusion inversion is a method that solves the inverse process of the aforementioned diffusion sampling equation. By introducing backward Euler iteration during the diffusion inversion process, we can obtain a more accurate solution.

\section{Proposed Method}
\label{sec:method}
The overall framework of the steganographic method proposed in this paper is shown in Figure \ref{fig:1}, which consists of three main components: message embedding, latent optimization, and backward Euler inversion.

\subsection{Message Embedding}
Given a string of message bits, the message embedding maps the message into an initial diffusion noise.
Let the latent space have dimensions $[F, T]$. We begin by initializing an orthogonal matrix $A \in \mathbb{R}^{N \times N}$, and define $C = F \times T \bmod (N \times N)$. The secret message is represented as a binary matrix $M \in \{0,1\}^{C \times N \times N}$. Using this, we compute the secret-embedded noise matrix as:
\begin{equation}
    z_{\mathrm{secret}} = A \cdot M \cdot A^T.
\end{equation}
Subsequently, $z_{\mathrm{secret}}$ is flattened into a vector, and additional noise sampled from a standard Gaussian distribution is appended to extend its length to $F \times T$. The resulting vector is then subjected to a shuffling operation and reshaped back into the $[F, T]$ latent space format, yielding the final noise tensor $z_s$ used for generation. This process can be succinctly expressed as:
\begin{equation}
    z_s = \mathrm{Reshape} \big( \mathrm{Shuffle} \big( \mathrm{Padding} ( A \cdot M \cdot A^T ) \big) \big).
\end{equation}
During extraction, the inverse operations are performed in reverse order: 
\begin{equation}
    M = A^T \cdot \mathrm{Clip} \big( \mathrm{Shuffle}^{-1} \big( \mathrm{Flatten}(z_s) \big) \big) \cdot A.
\end{equation}

\subsection{Latent Optimization}
Latent diffusion models employ an encoder-decoder architecture that compresses audio signals into latent representations, subsequently performing the diffusion generation process within the latent space. In steganographic applications, the encoder is utilized to transform steganographic audio into latent representations prior to inversion. However, due to the non-invertible nature of the encoder relative to the decoder, discrepancies arise between the encoder-reconstructed latent and the original latent representation, thereby impacting message extraction accuracy.

To address this issue, we employ neural network gradient optimization on the reconstructed latent, utilizing iterative approximation to converge toward the original latent representation prior to decoder computation. The detailed computational procedure is presented in Algorithm \ref{alg:1}.

\begin{algorithm}[!ht]
    \renewcommand{\algorithmicrequire}{\textbf{Input:}}
	\renewcommand{\algorithmicensure}{\textbf{Output:}}
	\caption{Power method}
    \label{alg:1}
    \begin{algorithmic}[1] 
        \REQUIRE  Encoder $E(\cdot)$, Decoder $D(\cdot)$, recived audio $x$, iteration steps $n$, iteration step size $h$;
	    \ENSURE Reconstructed latent $z$;
        
        \STATE $z \leftarrow E(x)$;
        
        \FOR {$i=1,2,\cdots,n$}
            \STATE $z = z - h \times \nabla_z\Vert x-D(z) \Vert^2_2$;
        \ENDFOR
   
        \STATE \textbf{return} $z$.
    \end{algorithmic}
\end{algorithm}

\subsection{Backward Euler Inversion}
Current initial noise embedding steganographic methodologies employ naive inversion methods for initial noise reconstruction. These approaches convert implicit inversion equations into explicit computational forms through approximation strategies, thereby enhancing computational efficiency while compromising inversion precision. 
Although such inversion methodologies are prevalent in diffusion-based editing applications, steganographic tasks necessitate prioritizing reconstruction accuracy over computational efficiency.

We introduce a more precise solving methodology by employing the backward Euler method to resolve implicit inversion equations, with detailed procedures provided for first-order and second-order solvers.

\subsubsection{First-Order Solver}
By rearranging Equation \ref{eq:1}, the implicit exact inversion equation can be obtained as follows:
\begin{equation}
    x_{t_{i-1}} = \frac{\sigma_{t_{i-1}}}{\sigma_{t_i}}\left( x_{t_i} + \alpha_{t_i}(e^{-h_i}-1)x_{\theta}(x_{t_{i-1}},t_{i-1})\right)
\end{equation}
Implicit equations cannot be computed directly but can be solved iteratively via the backward Euler method. The solution approach involves two steps: first, computing an initial approximation using the forward Euler method, and second, updating the implicit equation via Newton’s method or fixed-point iteration until convergence to the target accuracy is achieved. Let the iteration step size be $h$; the optimized DDIM inversion process is outlined in Algorithm 2.

\begin{algorithm}[!ht]
    \renewcommand{\algorithmicrequire}{\textbf{Input:}}
	\renewcommand{\algorithmicensure}{\textbf{Output:}}
	\caption{Backward Euler DDIM Inversion}
    \label{alg:2}
    \begin{algorithmic}[1] 
    \REQUIRE decoded latent $\hat{z}_{t_{N}}$, time steps $\{t_i\}_{i=0}^N$, data prediction model $z_{\theta}(\cdot)$, iteration stepsize $h$;
    \ENSURE initial noise $\hat{z}_{t_0}$;
    \FOR{$i\leftarrow N\;to\;1$}
        \STATE $\hat{z}_{t_{i-1}} \leftarrow \frac{\sigma_{t_{i-1}}}{\sigma_{t_i}}(\hat{z}_{t_i} + \alpha_{t_i}(e^{-h_i}-1)z_{\theta}(\hat{z}_{t_i},t_{i-1}))$     
        \WHILE{$|\hat{z}_{t_{i}}-z_{t_i}'|>\epsilon$}
            \STATE$z_{t_i}'\leftarrow \frac{\sigma_{t_i}}{\sigma_{t_{i-1}}}\hat{z}_{t_{i-1}}-\alpha_{t_i}(e^{-h_i}-1)z_{\theta}(\hat{z}_{t_{i-1}},t_{i-1})$
            \STATE$\hat{z}_{t_{i-1}}\leftarrow\hat{z}_{t_{i-1}}-h(z_{t_i}'-\hat{z}_{t_i})$
        \ENDWHILE
    \ENDFOR
    \STATE \textbf{return} $\hat{z}_{t_0}$.
    \end{algorithmic}
\end{algorithm}

Due to the unconditional stability of the backward Euler method, the convergence of the iteration is independent of the step size. Through backward Euler inversion, the error between adjacent time steps can be constrained within $\epsilon$.

\subsubsection{Second-Order Solver}
The proposed optimization method starts from the estimation of the second-order term and combines the forward Euler method (the conventional DDIM inversion) with the backward Euler method to balance accuracy and efficiency.

Consider Equation \ref{eq:2}. The errors introduced by the linear term and the first-order expansion term are relatively large, so we employ the backward Euler method for a more precise computation. In contrast, the approximation error introduced by the second-order term is relatively small; thus, we approximate it using the forward Euler method with a smaller step size and treat it as a constant during the backward Euler iterative process. The detailed procedure is described in Algorithm \ref{alg:3}.

\begin{algorithm}[!ht]
    \renewcommand{\algorithmicrequire}{\textbf{Input:}}
	\renewcommand{\algorithmicensure}{\textbf{Output:}}
	\caption{Second-order DPM-Solver Inversion}
    \label{alg:3}
    \begin{algorithmic}[1] 
    \REQUIRE Initial latent variable decoded $\hat{z}{t_N}$, sampling time steps ${t_i}^N{i=0}$, diffusion state update function $z_{\theta}(\cdot)$, backward Euler iteration update function UPDATE$(\cdot)$, sampling step size $J$;
    \ENSURE Reconstructed initial state $\hat{z}{t_0}$;
    \STATE $\lambda_{t_{i}} = \log\frac{\alpha_{t_{i}}}{\sigma_{t_{i}}},h_i=\lambda_{t_i}-\lambda_{t_{i-1}},r_i=\frac{h_{i-1}}{h_i}$
    \FOR{$i\leftarrow N\;to\;2$}
        \STATE $\hat{y}_{t_i}\leftarrow \hat{z}_{t_i}$
        \FOR{$j \leftarrow 1\;to\;2J$}
            \STATE $m_0=\alpha_{t_{i-j/J}}(e^{-h_{i-j/J}}-1)z_{\theta}(\hat{y}_{t_{i-(j-1)/J}},t_{i-j/J})$
            \STATE $\hat{y}_{t_{i-j/J}}\leftarrow\frac{\sigma_{t_{i-j/J}}}{\sigma_{t_{i-(j-1)/J}}}(\hat{y}_{t_{i-j/J}}+m_0)$
        \ENDFOR
        \STATE $\hat{z}_{t_{i-1}}\leftarrow \hat{y}_{t_{i-1}}$
        \WHILE{$|\hat{z}_{t_i}-z_{t_i}'|>\epsilon$}
            \STATE $m_0 = \alpha_{t_i}(e^{-h_i}-1)z_{\theta}(\hat{z}_{t_{i-1}},t_{i-1})$
            \STATE $m_1 = \frac{\alpha_{t_i}(e^{-h_i}-1)}{2r_i}(z_{\theta}(\hat{y}_{t_{i-1}},t_{i-1})-z_{\theta}(\hat{y}_{t_{i-2}},t_{i-2}))$
            \STATE $z_{t_i}'=\frac{\sigma_{t_i}}{\sigma_{t_{i-1}}}\hat{z}_{t_{i-1}}-m_0-m_1$
            \STATE$\hat{z}_{t_{i-1}}\leftarrow\hat{z}_{t_{i-1}}-h(z_{t_i}'-\hat{z}_{t_i})$
        \ENDWHILE
    \ENDFOR
    \STATE $\hat{z}_{t_0}\leftarrow\frac{\sigma_{t_0}}{\sigma_{t_1}}(\hat{z}_{t_1}+\alpha_{t_1}(e^{-h_i}-1)z_{\theta}(\hat{z}_{t_1},t_0))$
    \WHILE{$|\hat{z}_{t_1}-z_{t_1}'|>\epsilon$}
        \STATE $z_{t_1}'\leftarrow \frac{\sigma_{t_1}}{\sigma_{t_0}}\hat{z}_{t_0}-\alpha_{t_1}(e^{-h_1}-1)z_{\theta}(\hat{z}_{t_0},t_0)$
        \STATE$\hat{z}_{t_0}\leftarrow\hat{z}_{t_0}-h(z_{t_1}'-\hat{z}_{t_1})$
    \ENDWHILE 
    \STATE \textbf{return} $\hat{z}_{t_0}$
    
    \end{algorithmic}
\end{algorithm}

Since the forward sampling process of DPM-Solver initially samples the states required for higher-order sampling via lower-order forms, the backward Euler inversion is also divided into two stages. In Algorithm \ref{alg:3}, lines 16–20 correspond to the initial first-order sampling from $z_{t_0}$ to $z_{t_1}$, while lines 2–15 correspond to the subsequent second-order sampling process.

We analyze the algorithm as follows: the inversion of the second-order sampling process comprises two steps. The first step uses a fine-grained step-size forward Euler method to approximate the second-order term in Equation \ref{eq:2}, denoted by $\hat{y}_{t{i-1}}$ and $\hat{y}_{t{i-2}}$. The second step treats this second-order term approximation as a constant and applies the backward Euler method iteratively to the first-order term.

The fine-grained forward Euler computation is shown in lines 4–7. For illustration, a sampling step size of 50 is used, meaning DPM-Solver samples at 50 discrete time intervals. To obtain $\hat{y}_{t_{i-1}}$ and $\hat{y}_{t_{i-2}}$, this method does not directly perform traditional inversion over 50 time steps in one go. Instead, it iterates five times with a finer time step of 10 steps each. After obtaining $\hat{y}_{t_{i-1}}$, $\hat{y}_{t_{i-2}}$ is computed similarly from $\hat{y}_{t_{i-1}}$.

Lines 8–14 describe the backward Euler iterative process. The initial value for the iteration $\hat{z}_{t_{i-1}}$ is set as $\hat{y}_{t_{i-1}}$. In line 10, $m_0$ represents the first-order term of the Taylor expansion, while line 11’s $m_1$ corresponds to the second-order term, treated as constant here. The final state $\hat{z}_{t_{i-1}}$ is computed through the backward Euler iteration steps in lines 12 and 13.
Lines 16–20 implement the backward Euler process at time step $t_0$, equivalent to Algorithm \ref{alg:2}.

\subsection{Safety}
From a structural perspective, the carrier generation process in our method is similar to that proposed in \cite{hu2024establishing};
The only difference lies in the latent shape of diffusion models, thereby introducing padding and reshape methods.
\cite{hu2024establishing} provides a proof of distributional consistency, demonstrating that $z_s$ follows a standard Gaussian distribution, ensuring that the normal generation process and the steganographic process are indistinguishable, which ensures that the distribution and perceptual quality of stego audio are indistinguishable.

\begin{table*}[t]
\caption{Bit error rates(BER) of steganographic messages under different attack methods for each approach.(\%) $\downarrow$}\label{tab:1}
\begin{tabular}{ccccccccccccccc}
\hline
\multirow{2}{*}{Scheduler} & \multirow{2}{*}{method} & \multirow{2}{*}{ori} & \multicolumn{4}{c}{AAC compression(bps)} & \multicolumn{4}{c}{MP3 compression(bps)} & \multicolumn{2}{c}{resampling} & \multirow{2}{*}{\makecell{high\\atten}} & \multirow{2}{*}{\makecell{low\\boost}} \\ \cline{4-13}
                       &                         &                      & 320k    & 192K    & 128K   & 64K    & 320k    & 192K    & 128K   & 64K    & up             & down          &                             &                            \\ \hline
\multirow{4}{*}{DDIM}   & Yang\cite{yang2024gaussian}                   & 6.55                 & 6.81    & 6.81    & 6.81   & 9.57   & 6.65    & 6.65    & 6.74   & 8.31   & 6.55           & 7.20          & 7.23                        & 6.53                       \\
                       & Kim\cite{kim2025diffusion}               & 1.44                 & 1.54    & 1.54    & 1.54   & 2.58   & 1.46    & 1.47    & 1.50   & 2.22   & 1.43           & 1.74          & 2.42                        & 1.48                       \\
                       & Hu\cite{hu2024establishing}                     & 0.11                 & 0.12    & 0.12    & 0.13   & 0.26   & 0.12    & 0.11    & 0.12   & 0.19   & 0.11           & 0.13          & 0.46                        & 0.11                       \\
                       & PRoADS &\textbf{0.09}&\textbf{0.10}&\textbf{0.10}&\textbf{0.11}&\textbf{0.21}&\textbf{0.10}&\textbf{0.09}&\textbf{0.10}&\textbf{0.15}&\textbf{0.09}&\textbf{0.11}	&\textbf{0.25}&\textbf{0.09} \\ \hline
\multirow{4}{*}{DPMSolver}   & Yang\cite{yang2024gaussian}                   & 7.17                 & 7.44    & 7.44    & 7.45   & 9.09   & 7.27    & 7.27    & 7.33   & 8.68   & 7.16           & 7.74          & 7.69                        & 7.10                       \\
                       & Kim\cite{kim2025diffusion}                & 1.71                 & 1.80    & 1.81    & 1.82   & 2.97   & 1.78    & 1.80    & 1.82   & 2.56   & 1.70           & 2.09          & 2.44                        & 1.66                       \\
                       & Hu\cite{hu2024establishing}                    & 0.62                 & 0.64    & 0.63    & 0.62   & 0.84   & 0.64    & 0.63    & 0.63   & 0.75   & 0.62           & 0.67          & 0.83                        & 0.61                       \\
                       & PRoADS &\textbf{0.12}&\textbf{0.15}&\textbf{0.15}&\textbf{0.15}&\textbf{0.30}&\textbf{0.14}&\textbf{0.14}&\textbf{0.13}&\textbf{0.24}&\textbf{0.12}&\textbf{0.16}	&\textbf{0.29}&\textbf{0.13}	            \\ \hline
\end{tabular}
\end{table*}

\section{Experiment}
\label{sec:experiment}
\subsection{Setup}
During our preliminary experiments, we evaluated several publicly available audio diffusion models, including AudioLDM\cite{liu2023audioldm}, StableAudio\cite{evans2024fast}, MusicLDM\cite{chen2024musicldm}, and EzAudio\cite{hai2024ezaudio}. Ultimately, EzAudio was selected for its end-to-end trained Codec and relatively smaller parameter size. We conducted our experiments on the AudioCaps\cite{kim2019audiocaps} dataset. The model was employed to generate audio samples with a duration of 10 seconds and a sampling rate of 24 kHz. To ensure experimental fairness, a uniform embedding capacity of 57344($14\times64\times64$) bits was adopted for all steganographic systems.

To evaluate the robustness of our proposed method and baseline approaches under various attack scenarios, we introduce a comprehensive set of perturbations, including AAC and MP3 compression at four different bitrates, upsampling to 44.1 kHz, downsampling to 16 kHz, high-frequency attenuation (above 8 kHz, -6 dB), and low-frequency boost (below 200 Hz, +3 dB).

\subsection{Result}
\subsubsection{Robustness}
The BER of message extraction under various attack conditions for our proposed method and the comparison methods are presented in Table \ref{tab:1}. The message mapping algorithms of Yang\cite{yang2024gaussian} and Kim\cite{kim2025diffusion} exhibit high sensitivity to errors in initial noise reconstruction, resulting in consistently higher bit error rates across different attack scenarios.

Our method builds upon the mapping approach in Hu\cite{hu2024establishing} through three key modifications: an adjustment of the embedding algorithm to suit the latent space of varying shapes, coupled with two other techniques aimed at reducing extraction errors. Compared to Hu\cite{hu2024establishing}, under the DDIM sampling scheme, the bit error rate decreased by approximately 0.02\% for most attacks, with a reduction of 0.04\% observed under 64 kbps compression attacks and a more significant 0.21\% decrease under high-frequency attenuation attacks. Under the second-order DPMSolver sampling scheme, the bit error rate was reduced by about 0.5\% across various attacks.
These experimental results demonstrate that our proposed method effectively enhances the robustness of the steganographic system against a wide range of attack methods.

\begin{table}[h]
\caption{BER of ablation study.(\%) $\downarrow$}\label{tab:2}
\begin{tabular}{ccccc}
\hline
Attack & baseline & L.O. & B.E. & L.O.+B.E. \\
\hline
None                                                    & 0.62     & 0.22 & 0.19 & \textbf{0.12}      \\
AAC(192kbps)                                            & 0.63     & 0.23 & 0.22 & \textbf{0.15}      \\
AAC(64kbps)                                             & 0.84     & 0.39 & 0.39 & \textbf{0.30}      \\
down sampling                                           & 0.67     & 0.25 & 0.23 & \textbf{0.16}      \\
high atten                                              & 0.83     & 0.39 & 0.39 & \textbf{0.29}      \\
low boost                                               & 0.61     & 0.21 & 0.19 & \textbf{0.13}      \\
\hline
\end{tabular}
\end{table}

\subsubsection{Ablation}
To validate the effectiveness of Latent Optimization (L.O.) and Backward Euler Inversion (B.E.), ablation experiments were conducted. We remove the L.O. and B.E. methods to establish the baseline approach.
The comparison focused on several attack scenarios where robustness improvements were notably observed under the second-order DPMSolver sampling method. The results are presented in Table \ref{tab:2}.

For attacks with minor distortion impact (192 Kbps AAC compression, downsampling, and low-frequency enhancement), the B.E. method achieved better performance gains, reducing the BER by 0.01\%–0.03\%. Regarding 64 Kbps AAC compression and high-frequency attenuation attacks, both optimization methods exhibited comparable effectiveness. The synergistic application of both techniques yielded optimal results, delivering a 0.07\%–0.1\% BER reduction versus single-method implementations and approximately 0.5\% improvement over the baseline.

\subsection{Computation Cost}
The generation process of PRoADS is identical to the normal generation process. Producing a 10-second audio segment takes a total of 6.8 seconds, indicating that the sender can achieve continuous streaming generation of secret audio. For the inversion-based extraction process, however, due to the extensive iterations required to solve the inverse generation process, the time required is 106 seconds, which still falls within an acceptable range. Nevertheless, the improvements in reconstruction accuracy and bit error rate reduction are considered more critical in the context of steganography.

\section{Conclusion}
\label{sec:typestyle}
This paper proposes PRoADS, a provably secure and robust audio diffusion steganography scheme. Based on a robust orthogonal matrix projection embedding algorithm, by introducing Latent Optimization and Backward Euler Inversion, the scheme minimizes the reconstruction errors of initial noise and reduces the bit error rate of messages. Experiments demonstrate that our steganography scheme maintains a 0.3\% bit error rate even under 64 kbps AAC compression attacks.

\bibliographystyle{IEEEbib}
\bibliography{strings,refs}

\end{document}